%% file: main-ics-csr.tex
\documentclass{ewic}

\input{preamble}

\begin{document}

\runningheads{Barr\`ere 
    $\bullet$ Hankin
    $\bullet$ Eliades
    $\bullet$ Nicolaou
    $\bullet$ Parisini
}{Assessing Cyber-Physical Security in Industrial Control Systems}

\conference{
	6th International Symposium for ICS \& SCADA Cyber Security Research 2019 (ICS-CSR)\\{\normalfont\small \url{http://dx.doi.org/10.14236/ewic/icscsr19.7}}
} 

\title{Assessing Cyber-Physical Security in Industrial Control Systems}

\input{authors/authors-classic}

\newcommand{\content}{sections} 

\begin{abstract}
\input{\content/abstract}
\end{abstract}


\maketitle

\input{\content/intro}
\input{\content/background}
\input{\content/multiple-measures}

\input{\content/model-strategy}
\input{\content/analytical-evaluation}

\input{\content/case-study}
\input{\content/related-work}
\input{\content/conclusions}

\section*{ACKNOWLEDGMENTS}
We would like to thank the reviewers for their helpful and valuable comments. 
\ackContent

\input{biblio}
\end{document}

%% file: preamble.tex
\pdfoutput=1
\usepackage{hyperref}
\hypersetup{
    colorlinks=true,
    linkcolor=black,
    citecolor=black,
    filecolor=black,
    urlcolor=black,
}

\usepackage{graphicx}
\usepackage{subfig}
\usepackage{wrapfig}

\usepackage[utf8]{inputenc}
\usepackage[english]{babel}
\usepackage{verbatim} 
\usepackage{amsmath,amssymb} 
\usepackage{listings} 
\usepackage{color}
\usepackage[dvipsnames]{xcolor}
\usepackage{url} 

\usepackage{graphicx}

\usepackage[oldcommands,linesnumbered,boxed]{algorithm2e}
\setalcapskip{1em} 

\newcommand{\martinAlgFontsize}{\fontsize{8.0}{8.0}\selectfont}
\SetAlFnt{\martinAlgFontsize}
\newcommand{\martinAlgCapFontsize}{\fontsize{9.0}{9.0}\selectfont}
\SetAlCapNameFnt{\martinAlgCapFontsize}
\SetAlCapFnt{\martinAlgCapFontsize}

\newcommand{\martinTableFontsize}{\scriptsize\selectfont}

\newcommand{\martinListingFontsize}{\fontsize{8}{8}\selectfont}

\SetKwInOut{Global}{Global}
\SetKwInOut{Name}{Name}

\newcommand{\ackContent}{
This work has been supported by the European Union's Horizon 2020 research and innovation programme under grant agreement No. 739551 (KIOS CoE).
}

\newcommand{\andnodes}{\Delta} \newcommand{\ornodes}{\Theta}

\newcommand{\gdomain}{\mathcal{G}}
 
\newcommand{\cost}{\varphi} 
\newcommand{\costprime}{\psi} 
\newcommand{\form}{f_{G}} 
\newcommand{\expform}{h_{G}}

\newcommand{\wcc}{wcc} 
\newcommand{\rem}{\sigma} 
\newcommand{\tool}{META4ICS } 
\newcommand{\toolx}{META4ICS}

\colorlet{punct}{red!60!black}
\definecolor{background}{HTML}{FFFFFF}
\definecolor{delim}{RGB}{20,105,176}
\colorlet{numb}{magenta!60!black}

\lstdefinelanguage{json}{
	basicstyle=\normalfont\ttfamily,
	numberstyle=\scriptsize,
	stepnumber=1,
	numbersep=8pt,
	showstringspaces=false,
	breaklines=true,
	frame=lines,
	stringstyle=\color{blue},
	backgroundcolor=\color{background},
	literate=
	*{0}{{{\color{numb}0}}}{1}
	{1}{{{\color{numb}1}}}{1}
	{2}{{{\color{numb}2}}}{1}
	{3}{{{\color{numb}3}}}{1}
	{4}{{{\color{numb}4}}}{1}
	{5}{{{\color{numb}5}}}{1}
	{6}{{{\color{numb}6}}}{1}
	{7}{{{\color{numb}7}}}{1}
	{8}{{{\color{numb}8}}}{1}
	{9}{{{\color{numb}9}}}{1}
	{:}{{{\color{punct}{:}}}}{1}
	{,}{{{\color{punct}{,}}}}{1}
	{\{}{{{\color{delim}{\{}}}}{1}
	{\}}{{{\color{delim}{\}}}}}{1}
	{[}{{{\color{delim}{[}}}}{1}
	{]}{{{\color{delim}{]}}}}{1},
}

\definecolor{eclipseStrings}{RGB}{42,0.0,255}
\definecolor{eclipseKeywords}{RGB}{127,0,85}
\colorlet{numb}{magenta!60!black}

\lstdefinelanguage{json2}{
	basicstyle=\normalfont\ttfamily,
	commentstyle=\color{eclipseStrings}, 
	stringstyle=\color{eclipseKeywords}, 
	numbers=left,
	numberstyle=\scriptsize,
	stepnumber=1,
	numbersep=8pt,
	showstringspaces=false,
	breaklines=true,
	frame=lines,
	backgroundcolor=\color{white}, 
	string=[s]{"}{"},
	comment=[l]{:\ "},
	morecomment=[l]{:"},
	literate=
	*{0}{{{\color{numb}0}}}{1}
	{1}{{{\color{numb}1}}}{1}
	{2}{{{\color{numb}2}}}{1}
	{3}{{{\color{numb}3}}}{1}
	{4}{{{\color{numb}4}}}{1}
	{5}{{{\color{numb}5}}}{1}
	{6}{{{\color{numb}6}}}{1}
	{7}{{{\color{numb}7}}}{1}
	{8}{{{\color{numb}8}}}{1}
	{9}{{{\color{numb}9}}}{1}
}

%% file: authors/authors-classic.tex
\authorone{Mart\'in Barr\`ere, Chris Hankin\\
    Institute for Security Science and Technology\\
    Imperial College London, UK\\
    \email{\{m.barrere, c.hankin\}@imperial.ac.uk}}

\authortwo{Demetrios G. Eliades, Nicolas Nicolaou\\
    KIOS Research and Innovation Centre of Excellence\\
    University of Cyprus\\    
    \email{\{eldemet, nicolasn\}@ucy.ac.cy}}

\authorthree{Thomas Parisini\\
   Department of Electrical and Electronic Engineering\\ 
   Imperial College London, UK\\
    \email{t.parisini@imperial.ac.uk}}

%% file: sections/abstract.tex
Over the last years, Industrial Control Systems (ICS) have become increasingly exposed to a wide range of cyber-physical threats. Efficient models and techniques able to capture their complex structure and identify critical cyber-physical components are therefore essential. AND/OR graphs have proven very useful in this context as they are able to semantically grasp intricate logical interdependencies among ICS components. However, identifying critical nodes in AND/OR graphs is an NP-complete problem. In addition, ICS settings normally involve various cyber and physical security measures that simultaneously protect multiple ICS components in overlapping manners, which makes this problem even harder. In this paper, we present an extended security metric based on AND/OR hypergraphs which efficiently identifies the set of critical ICS components and security measures that should be compromised, with minimum cost (effort) for an attacker, in order to disrupt the operation of vital ICS assets. Our approach relies on MAX-SAT techniques, which we have incorporated in \toolx, a Java-based security metric analyser for ICS. We also provide a thorough performance evaluation that shows the feasibility of our method. Finally, we illustrate our methodology through a case study in which we analyse the security posture of a realistic Water Transport Network~(WTN). 

%% file: sections/intro.tex
\section{Introduction}
\label{sec:intro}

For many decades, Industrial Control Systems~(ICS) such as water treatment plants, energy, oil, gas plants, and others, have been safely operated in isolation from the external world. 
However, with the advent of the Internet, new convenient IT-based control mechanisms, and highly interconnected networks, ICS environments have become an appealing target for malicious actors. 
Cyber attacks on these systems can have devastating consequences such as flooding, blackouts, or even nuclear disasters (\cite{Humayed2017}). 
Stuxnet, Industroyer, NotPetya, and more recently, WannaCry, exemplify the impact this type of attack may have on critical ICS infrastructures (\cite{CyberXReport2019};  \cite{Lee2016}; \cite{Falliere2011}). 
Therefore, protecting industrial control systems from cyber threats is a high priority as their compromise can result in a myriad of different problems, from service disruptions and economical loss, to jeopardising natural ecosystems and putting human lives at risk. 

Due to the complex nature of cyber-physical systems and the convoluted web of dependencies among their components, it is paramount to count with appropriate models and tools able to measure ICS security and prioritise their weakest points. 
In particular, identifying critical ICS nodes not only allows to understand the security level of a system but also provides actionable information that can be used to decide how and where to improve security as well as adding redundant and fallback components to increase reliability. 
While previous works do propose different techniques to quantify ICS security levels, they usually consider an individual security score for each ICS component that are then combined based on the underlying logical connectivity, e.g. AND/OR dependency graphs (\cite{BarrereSecMetric2019}). 
However, when various cyber-physical security measures (often disregarded in physical environments) are applied simultaneously to protect one or more ICS components altogether (e.g. fenced areas, alarm systems, authentication procedures), the use of independent scores on each ICS component might fail to capture the overall security level of the system. 

In this paper, we present a novel approach based on AND/OR hypergraphs that is able to efficiently identify the set of critical components and security measures, with the lowest compromise cost (effort) for an attacker, whose violation would imply an operational disruption to the ICS system. 
Our approach builds upon the model presented in (\cite{BarrereSecMetric2019}) and extends the strategy to address multiple overlapping security measures. 

\textbf{Our main contributions are: }
(1) a mathematical model able to represent multiple overlapping security measures over complex AND/OR dependency graphs for ICS environments, 
(2) an efficient security metric to identify critical cyber-physical components and security measures, 
(3) an implementation prototype based on META4ICS (\cite{BarrereMeta4icsGithub}), 
(4) an extensive experimental evaluation on performance and scalability aspects, and   
(5) a case study conducted on a realistic water transport network that shows the applicability of our security metric. 

%% file: sections/background.tex
\vspace{-0.3cm}
\section{Background concepts}
\label{sec:background}
In this section, we recall the main concepts of the base security metric (\cite{BarrereSecMetric2019}). 

\subsection{Network graph modelling}

An industrial network $W$ is modelled as an AND/OR graph $G = (V,E)$ that represents the operational dependencies in $W$. The graph involves three types of basic vertices, called atomic nodes ($V_{AT}$), that model different network components: $S$ represents the set of sensor nodes, $C$ represents the set of actuator nodes, and $A$ represents the set of software agents (running for example in PLCs and RTUs). 
We define $V_{AT} =  S \cup C \cup A$. In addition, the graph also involves two artificial node types that model logical dependencies between network components: $\andnodes$ represents the set of logical AND nodes, and $\ornodes$ represents the set of logical OR nodes. The set of all graph nodes is defined as  $V(G) = V_{AT} \cup \andnodes \cup  \ornodes$. 

$E(G)$ corresponds to the set of edges among nodes and their semantics depend on the type of nodes they connect.
Roughly stated, an edge $(a_1,a_2)$ means that node $a_2$ depends on $a_1$ to work properly. 
The graph also involves AND and OR nodes, which act as special connectors and are interpreted from a logical perspective. If a node $v$ is reached by an OR node, this means that the operational purpose of $v$ can be satisfied, i.e. $v$ operates normally, if \textit{at least one} of the incoming nodes to the OR node is also satisfied. 
Alike, a node $w$ reached by an AND node will be satisfied if \textit{all} of the incoming nodes to the AND node are also satisfied. 

\subsection{Simple example}
\begin{figure}[!h]
    \centering
    \includegraphics[scale=0.25]{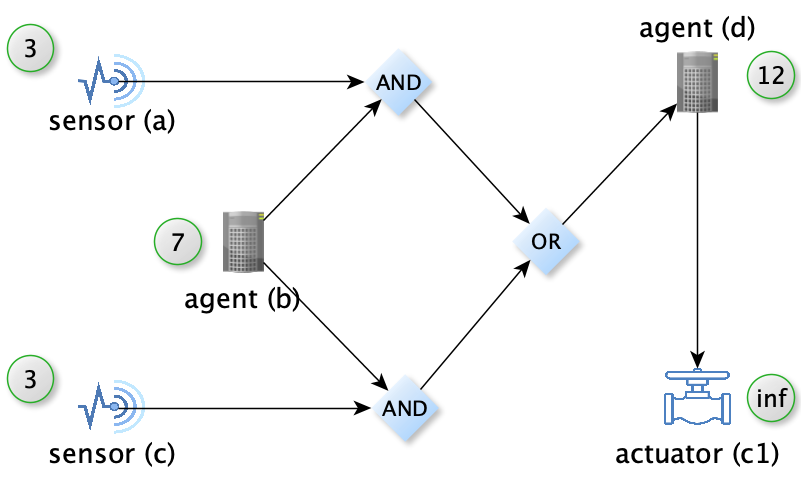}
    \caption{Case 1 - Individual protection measures}
    \label{fig:case1}
\end{figure}

Let us consider the scenario illustrated in Figure \ref{fig:case1}. 

In this case, the AND/OR graph reads as follows: the actuator $c1$ depends on the output of software agent $d$. Agent $d$, in turn, has two alternatives to work properly; it can use either the readings of sensor $a$ and the output from agent $b$ together, or the output from agent $b$ and the readings of sensor $c$ together. 
Focused on actuator $c1$, the metric models these dependencies from a logical perspective as follows: 
\begin{lstlisting}[
label={lst:example1-exec1}, 
mathescape=true, language=bash, %caption=Security metric resolution, 
basicstyle=\martinListingFontsize\sffamily, 
keywordstyle=\bfseries\color{green!40!black}, 
commentstyle=\itshape\color{purple!40!black},
identifierstyle=\color{blue},
stringstyle=\color{orange}, 
frame=single, 
%numbers=left,
numberstyle=\tiny\color{gray},
stepnumber=1, 
belowcaptionskip=5em,
belowskip=3em, 
captionpos=b, 
%xleftmargin=.125\textwidth, xrightmargin=.125\textwidth
]                   
          $\form(c1) = c1 \land d \land ( (a \land b) \lor (b \land c) ) $
\end{lstlisting}

\vspace{-1.0cm}
More formally, $\form(t)$ produces a propositional formula that represents the logical semantics of $G$ with regards to $t$, i.e. the logical conditions (dependencies) that must be satisfied to fulfil node~$t$. In addition, each CPS component has associated an individual score that represents its compromise cost where \textit{inf} means infinite. 

The objective of the attacker is to falsify $\form(c1)$, thus satisfy the formula $\neg\form(c1)$, in order to make $c1$ unable to operate. The metric considers the compromise costs of the nodes as the weights of the logical variables, in the form of a cost function $\cost(n) : V_{AT} \rightarrow \mathbb{R}_{\geq0}$, and then builds a Weighted Partial MAX-SAT problem. A solution to this problem indicates the nodes that should be compromised in order to disrupt the operation of actuator c1, with minimal effort (cost) for the attacker.  

In our example, the least-effort attack strategy to disable actuator $c1$ is $\{a,c\}$ with a total cost of $6$. 
The compromise of sensors $a$ and $c$ will disable both AND nodes, and consecutively the OR node, 
which in turn will affect node $d$ and finally node $c1$. 
We formalise the original security metric in the next section.

\subsection {Base security metric definition} 

Let $W$ be an industrial network, $G = (V,E)$ a directed AND/OR graph representing the operational dependencies in $W$, and $t$ a target network node. 
The objective of the security metric, denoted as $\mu(G,t)$, is to identify the set of nodes $X=\{x_1, \ldots, x_h\}$ that must be compromised in order to disrupt the normal operation of target node $t$, with minimal cost for the attacker. 
More formally, $\mu : \gdomain \times V \rightarrow 2^{V}$ is defined as follows: 
\begin{equation}
\begin{array}{c}
\displaystyle \mu(G,t) = \underset{X \subseteq V_{AT}}{\text{argmin} } \Big( \sum_{x_i \in X} \cost(x_i) \Big) \\ 
\textrm{s.t.}\\
\displaystyle \wcc(\rem(G, X)) \geq 2 \lor X = \{t\}\\
\end{array}
\end{equation}
where the solution with minimal cost must be either node $t$ or a set of nodes $X$ such that, if removed (with function $\rem$), $t$ gets disconnected from the graph. 
Function $\rem (G,X)$ removes from $G$ each node $x \in X$ and the nodes that depend on them following a logic-style propagation, as explained in~\cite{BarrereSecMetric2019}. 
The result is then analysed with function $\wcc(G)$, which computes the number of weakly connected components in $G$, that is, the number of connected components when the orientation of edges in $G$ is ignored. In other words, the restriction on $\wcc(G)$ ensures that the target node $t$ is disconnected from a non-empty set of nodes on which $t$ depends (directly or indirectly) to function properly. 

%% file: sections/multiple-measures.tex
\vspace{-0.1cm}
\section{Using multiple overlapping security measures}
\label{sec:overlapping}

\begin{figure}[!t]
    \centering
    \includegraphics[scale=0.22]{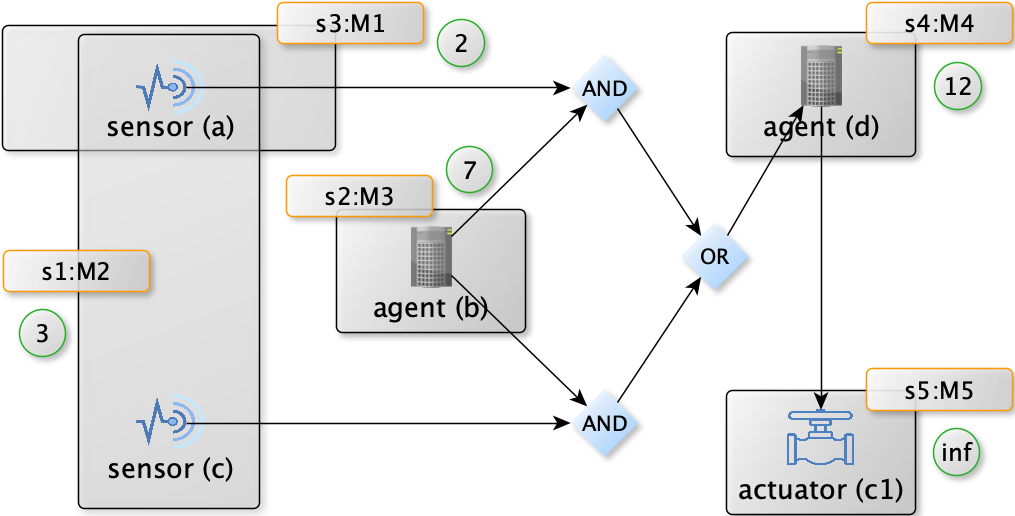}
    \caption{Case 2 - Multiple overlapping measures}
    \vspace{-0.4cm}
    \label{fig:case3}
\end{figure}

While quite useful, the original metric is only able to capture cyber-physical security measures that are applied independently to each ICS component. That is, it can capture that sensors $a$ and $c$ are protected, for example, by fenced areas (each one with cost $3$), but it cannot model that both sensors are protected by one single fenced area with cost $3$. In other words, the metric assumes that these two fenced areas are different, and thus compromising one of them does not affect the other. In mathematical terms, this means that the costs for the attacker are completely independent. Nevertheless, if the fenced area is the same for both sensors, then the attacker's effort (cost) required to compromise the security mechanism must be considered only once. 
Let us consider a second example, illustrated in Figure \ref{fig:case3}. 

{
    \renewcommand{\arraystretch}{1.35}
    \begin{table}[!b]
        \martinListingFontsize
        \centering
        \begin{tabular}{|c|c|c|}
            \hline
            {Measure} & {Cost (attacker)} & {Description}\\
            \hline
            {$M1$} & {$2$} & {Sound alarm} \\
            \hline          
            {$M2$} & {$3$} & {Fenced area} \\
            \hline
            {$M3$} & {$7$} & {Locked container} \\
            \hline
            {$M4$} & {$12$} & {Tamper-resistant container } \\
            \hline            
            {$M5$} & {\textit{inf}} & {Alarmed locked building} \\
            \hline            
        \end{tabular}
        \vspace{0.2cm}
        \caption{Protection measures}
        \label{tab:measures-examples-all}
        \vspace{-0.36cm}
    \end{table}
}

This second scenario describes a more general problem where many security measures, as those exemplified in Table~\ref{tab:measures-examples-all}, can be jointly applied to protect multiple ICS components simultaneously. 

In particular, each ICS component is protected by one or more security measure instances $s_j$ of type $M_i$, as described in Table \ref{tab:measures-case3}. We define $S=\{s1, s2, \ldots\}$ as the set of involved security measure instances. We call \textit{protection range} to the set of ICS components protected by a single instance $s_j$. 

\vspace{-0.1cm}
{
	\renewcommand{\arraystretch}{1.35} 
	\begin{table}[!h]
         \martinListingFontsize
		\centering
		\begin{tabular}{|c|c|c|c|c|c|}
			\hline
			{Measure instance} & {$s1$} & {$s2$} & {$s3$} & {$s4$} & {$s5$} \\
			\hline
			{Measure type} & {$M2$} & {$M3$} & {$M1$} & {$M4$} & {$M5$}\\
			\hline
			{Attacker's cost  $\costprime(s_j)$} & {$3$} & {$7$} & {$2$} & {$12$} & {$inf$}\\
			\hline
			{Protection range} & {$\{a,c\}$} & {$\{b\}$} & {$\{a\}$} & {$\{d\}$} & {$\{c1\}$}\\
			\hline
		\end{tabular}
		\vspace{0.2cm}
		\caption{Security measures for Case 2}
		\label{tab:measures-case3}
	\end{table}
}
\vspace{-0.5cm}

Each measure $M_i$ involves a cost for the attacker that quantifies the effort that he or she has to make in order to bypass the measure. We model this aspect for measure instances as a function $\costprime: S \rightarrow \mathbb{R}_{\geq0}$. 

In the second scenario, sensors $a$ and $c$ are protected by the same security measure instance $s1$ (fenced area). Therefore, the cost of bypassing $s1$ to compromise sensor $a$, sensor $c$, or both, is $3$. However, sensor $a$ is also protected by the security measure $s3$ (sound alarm). As a consequence, compromising sensor $a$ would imply to bypass both protective measures $s1$ and $s3$. Therefore, the best strategy in this case is~to compromise the security measures $s1$ and $s3$, involving the critical nodes $a$ and $c$, with a total cost of $3+2=5$. Note that the original metric would have counted $3 + 2$ for sensor $a$ and $3$ for sensor $c$, totalling a cost of $8$. 

In the next section, we formalise an hypergraph-based extension to the base metric described in Section~\ref{sec:background} that allows to capture multiple security measures applied to various ICS components in overlapping manners.

%% file: sections/model-strategy.tex
\section{Extended security metric}
\label{sec:resolution-strategy}

\subsection{Mathematical reformulation}

We redefine the security metric $\mu(G,t)$ as follows: 
\begin{equation}
    \label{eq:extende_metric}
\hspace{-0.2cm}
\begin{array}{c}
\displaystyle \mu(G,t) = \underset{X \subseteq V_{AT}}{\text{argmin} } \Big(
\sum_{x_i \in X} \cost(x_i) + \sum_{s_j \in S(X)} \costprime(s_j)  
\Big) \\ 
\textrm{s.t.}\\
\displaystyle \wcc(\rem(G, X)) \geq 2 \lor X = \{t\}\\
\end{array}
\end{equation}

where function $S(X)$ returns the set of security measure instances used to protect the nodes in~$X$. Since  $S(X)$ returns a set, measure instances that protect more than one node  in $X$ appear only once, and thus their costs are considered only once in Equation~\ref{eq:extende_metric}. Note also that  $\cost(n)$ can be neutral (e.g. $\cost(n)=0, \forall n \in V_{AT}$) to only consider the costs of the security measures, or it can be instantiated with cyber costs, e.g. CVSS scores (\cite{CVSS}).

\subsection{AND/OR hypergraph formalisation}

Hypergraphs are a generalisation of standard graphs where graph edges, called hyperedges, can connect any number of vertices (\cite{Berge1989}). More formally, let $X$ be a set of vertices $X=\{x_1,x_2,\ldots, x_n\}$. A hypergraph on $X$, denoted $H = (X,E)$, is a family of subsets of $X$, with $E = \{e_1,e_2,\ldots,e_m\}$, such that: (1) there are no empty edges in $H$, i.e. $e_i \neq \emptyset, \forall e_i \in E$; and (2) $X$ is covered by $E$, i.e. $\bigcup_{i=1}^m e_i = X$.

In this work, we propose the use of a hybrid type of hypergraph, called AND/OR hypergraph, which essentially combines properties of hypergraphs and the logical structure of AND/OR graphs. Roughly stated, the nodes of an AND/OR hypergraph are the hyperedges of a standard hypergraph, and these are linked using logical AND/OR nodes as done in classical AND/OR graphs. 

We use standard hypergraphs to model groups of security measures that are applied to each ICS component in the network. For example, let us consider Case 2 illustrated in Figure \ref{fig:case3}. In this case, the hypergraph is defined as $H=(X,E)$ where $X=V_{AT} \cup S$ is the set of nodes of the hypergraph, and $E=\{e_1,e_2,e_3,e_4,e_5\}$ is the set of hyperedges. Table \ref{tab:hypergraph-case3} details the members of each hyperedge $e_i \in E$. 

{
	\renewcommand{\arraystretch}{1.35}
	\begin{table}[!h]
        \martinListingFontsize
		\centering
		\begin{tabular}{|c|c|c|c|c|}
			\hline
			{$e_1$} & {$e_2$} & {$e_3$} & {$e_4$} & {$e_5$} \\
			\hline
			{$\{a, s1, s3\}$} & {$\{c,s1\}$} & {$\{b,s2\}$} & {$\{d,s4\}$} & {$\{c1,s5\}$} \\
			\hline
		\end{tabular}
		\vspace{0.2cm}
		\caption{Hypergraph $H$ for case 3}
		\label{tab:hypergraph-case3}
	\end{table}
}
\vspace{-0.6cm}
Hyperedges combine each network node with the instances of the security measures that are used to protect them. The advantage of using hypergraphs is that we can capture multiple overlapping security measures in the hyperedges of the hypergraph. In addition, we can easily model protection ranges, that is, how a specific measure instance, e.g. a fenced area, protects multiple ICS components simultaneously, e.g. $s1 \mapsto \{a,c\}$. 

At a semantic level, the interpretation of a hyperedge $e_i$ is that the original node $n$ is accompanied by the security measures that protect it, and therefore, node $n$ can only be disrupted if every security measure in $e_i$ is compromised too. Now hyperedges can be understood as super nodes that represent each original node and their protective measures. Therefore, we can follow the same logical structure as in the original graph and combine these super nodes via AND/OR connectives as illustrated in Figure \ref{fig:and-or-graph-hyperedges-case3}. From a logical perspective, we map the dependency model of the AND/OR hypergraph as follows: 

\begin{figure}[!t]
	\centering
	\includegraphics[scale=0.21]{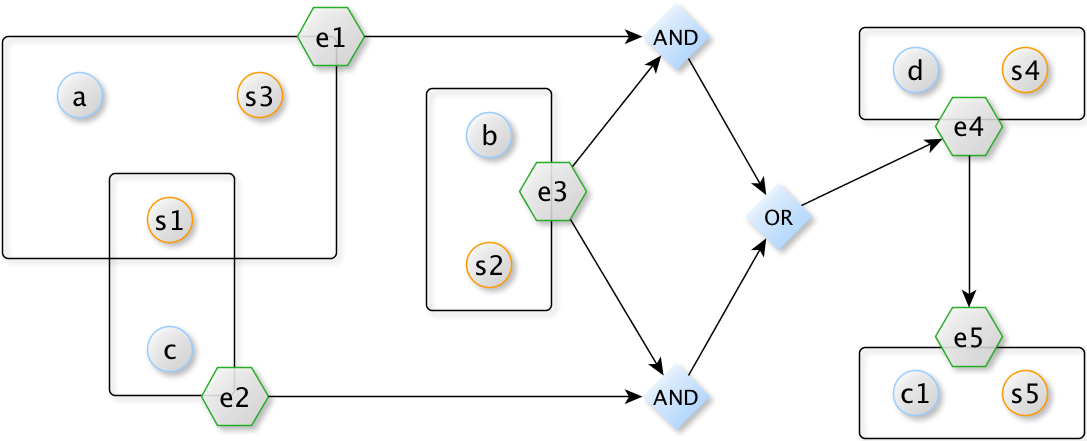}
	\caption{AND/OR hypergraph for Case 2}
	\vspace{-0.5cm}
	\label{fig:and-or-graph-hyperedges-case3}
\end{figure}

\begin{lstlisting}[
label={lst:example1-exec1}, 
mathescape=true, language=bash, %caption=Security metric resolution, 
basicstyle=\martinListingFontsize\sffamily, 
keywordstyle=\bfseries\color{green!40!black}, 
commentstyle=\itshape\color{purple!40!black},
identifierstyle=\color{blue},
stringstyle=\color{orange}, 
frame=single, 
%numbers=left,
numberstyle=\tiny\color{gray},
stepnumber=1, 
belowcaptionskip=5em,
belowskip=3em, 
captionpos=b, 
%xleftmargin=.125\textwidth, xrightmargin=.125\textwidth
]                   
        $f_H(e5) = e5 \land e4 \land ( ( e1 \land e3 ) \lor ( e3 \land e2 ) ) $
\end{lstlisting}
\vspace{-0.8cm}

As explained in (\cite{BarrereSecMetric2019}), the objective of the attacker is to falsify the previous formula (or satisfy  $\neg f_H(e5)$) in order to make the target $e5$ non-functional. Since each hyperedge $e_i$ involves many security measures plus the original node $n$, the only way to falsify $e_i$ is to falsify every member in it. Therefore, we logically capture this aspect by replacing each hyperedge $e_i$ by a disjunctive construct $(n \lor s_i \lor \ldots \lor s_j)$, where $s_i \lor \ldots \lor s_j$ is the disjunction of measure instances that protect node $n$. Such a disjunctive construct actually forces a SAT solver to make every security measure \textit{false}, which essentially equals to the fact that the attacker must compromise all of the measures to take control of the ICS component. 

Considering the costs of the security measures as weights for the logical variables, we extend the \hbox{MAX-SAT} problem specification as explained in the following section.

\subsection{Weighted Partial MAX-SAT problem specification}
\label{sec:problem-spec}

The following steps describe the actions required to prepare the specification of the MAX-SAT problem. 

\begin{enumerate}[topsep=-5pt,itemsep=-1ex,partopsep=0ex,parsep=1ex]
	\item Traverse the dependency graph $G$ and build an equivalent logical representation, $\form(t)$, as explained in Section \ref{sec:background}. 
	\vspace{0.2cm}
	
	\item Build a new formula $\expform(t)$ by replacing each atomic node $n \in V_{AT}$ in $\form(t)$ with $(n \lor s_i \lor \ldots \lor s_j)$, where $s_i \lor \ldots \lor s_j$ is the disjunction of security controls that protect node $n$. 
	\vspace{0.2cm}	
	
	\item Transform the attacker's objective $\neg\expform(t)$ into an equisatisfiable CNF formula using the Tseitin transformation (\cite{Tseitin70}). 
	\vspace{0.2cm}	
	
    \item Consider $\costprime(s_i)$ as the penalty cost of each variable $s_i$ and $\cost(n)$ for atomic nodes. 
	\vspace{0.2cm}	
		
\end{enumerate}

Finally, the Weighted Partial MAX-SAT problem is instantiated as $\neg\expform(t)$, which is the objective of the attacker,  and solved by \tool (\cite{BarrereMeta4icsGithub}) as described in~(\cite{BarrereSecMetric2019}).

\subsection{Execution example over Case 2}
Let us reconsider Case 2 illustrated in Figure~\ref{fig:case3}. This scenario can be logically formulated as follows: 
\begin{lstlisting}[
label={lst:example1-exec1}, 
mathescape=true, language=bash, %caption=Security metric resolution, 
basicstyle=\martinListingFontsize\sffamily, 
keywordstyle=\bfseries\color{green!40!black}, 
commentstyle=\itshape\color{purple!40!black},
identifierstyle=\color{blue},
stringstyle=\color{orange}, 
frame=single, 
%numbers=left,
numberstyle=\tiny\color{gray},
stepnumber=1, 
belowcaptionskip=5em,
belowskip=3em, 
captionpos=b, 
%xleftmargin=.125\textwidth, xrightmargin=.125\textwidth
]                   
          $\form(c1) = c1 \land d \land ( (a \land b) \lor (b \land c) ) $
\end{lstlisting}

\vspace{-0.9cm}
Based on the protective measures, the AND/OR hypergraph is logically mapped as follows: 

\begin{lstlisting}[
label={lst:example1-exec1}, 
mathescape=true, language=bash, %caption=Security metric resolution, 
basicstyle=\martinListingFontsize\sffamily, 
keywordstyle=\bfseries\color{green!40!black}, 
commentstyle=\itshape\color{purple!40!black},
identifierstyle=\color{blue},
stringstyle=\color{orange}, 
frame=single, 
%numbers=left,
numberstyle=\tiny\color{gray},
stepnumber=1, 
belowcaptionskip=5em,
belowskip=3em, 
captionpos=b, 
%xleftmargin=.125\textwidth, xrightmargin=.125\textwidth
]                   
        $f_H(e5) = e5 \land e4 \land ( ( e1 \land e3 ) \lor ( e3 \land e2 ) ) $
\end{lstlisting}

\vspace{-1.1cm}
The new formulation $\expform(t)$ produced at step 2 is as follows: 
\begin{lstlisting}[
label={lst:example1-exec1}, 
mathescape=true, language=bash, %caption=Security metric resolution, 
basicstyle=\martinListingFontsize\sffamily, 
keywordstyle=\bfseries\color{green!40!black}, 
commentstyle=\itshape\color{purple!40!black},
identifierstyle=\color{blue},
stringstyle=\color{orange}, 
frame=single, 
%numbers=left,
numberstyle=\tiny\color{gray},
stepnumber=1, 
belowcaptionskip=5em,
belowskip=3em, 
captionpos=b, 
%xleftmargin=.125\textwidth, xrightmargin=.125\textwidth
]                   
 $\expform(c1) = (c1 \lor s5) \land (d \lor s4) \;\land $ 
        $(  ( (a \lor s1 \lor s3) \land (b \lor s2)) \lor ( (b \lor s2) \land (c \lor s1)) ) $
\end{lstlisting}

\vspace{-0.9cm}
If we now consider, for example, a unit cost on each atomic node $n$, i.e. $\cost(n)=1, \forall n \in V_{AT}$, the solution of the Weighted Partial MAX-SAT problem for $\neg\expform(c1)$ is composed of instances $s1$ and $s3$ with a total cost of $7$. Informally speaking, we are trying to find a portion of $\expform(c1)$ that can be falsified (so $\neg\expform(c1)$ is \textit{true}) with minimal cost. Table \ref{tab:strategy-example-measures} shows the attacker's costs for each measure instance that are used as the falsification penalty scores. 
{
	\renewcommand{\arraystretch}{1.35}
	\begin{table}[!h]
         \martinListingFontsize
		\centering
		\begin{tabular}{|c|c|c|c|c|c|}
			\hline
			{Measure instance} & {$s1$} & {$s2$} & {$s3$} & {$s4$} & {$s5$} \\
			\hline
			{Cost (attacker)} & {$3$} & {$7$} & {$2$} & {$12$} & {$inf$}\\
			\hline
		\end{tabular}
		\vspace{0.2cm}
		\caption{Falsification penalty scores}
		\label{tab:strategy-example-measures}
	\end{table}
}

\vspace{-0.9cm}
We can observe that if the last big clause of $\expform(c1)$ (line 2) is falsified, then $\expform(c1)$ is falsified. We can choose to falsify the whole disjunction by making, for example, the sub-sentence $(b \lor s2)$ \textit{false}. However, the penalty here is $1 + 7 = 8$. If $(a \lor s1 \lor s3)$ and $(c \lor s1)$ are falsified instead, the cost corresponds to the penalty paid for the set $\{a,s1,s3,c\}$ with a total cost of $1 + 3 + 2 + 1=7$. The other two options, $(c1 \lor s5)$ and $(d \lor s4)$, have costs \textit{infinite} and $13$ respectively, so the final solution involves the critical node set $\{a,c\}$ and measures $\{s1,s3\}$ with a total cost of $7$. 

\vspace{-0.2cm} 

%% file: sections/analytical-evaluation.tex
\section{Performance evaluation}
\label{sec:analytical-exp}

We have performed a thorough experimental analysis that shows the feasibility and performance of our approach. In this section, we first describe the tool and methods used within the experiments. Afterwards, we explain the obtained results for independent security measures applied across the graph. Finally, we study the use of various security measures applied to multiple nodes simultaneously and the impact this overlapping poses in terms of computation time. The experimental evaluation has been performed using a MacBook Pro (15-inch, 2018), 2.9 GHz Intel Core i9, 32 GB 2400 MHz~DDR4. 

\subsection{Implementation prototype and AND/OR graph generation}

Our implementation prototype relies on \tool (\cite{BarrereSecMetric2019}), a Java-based security metric analyser for ICS, available at (\cite{BarrereMeta4icsGithub}). \tool consumes JSON specification files that describe ICS environments in the form of AND/OR graphs, and outputs their security score as well as the critical nodes that require utmost attention. In this work, we have extended \tool in order to cover hypergraph-related concepts and the application of multiple overlapping security measures over ICS network components.  

Within our experiments, we use synthetic pseudo-random AND/OR graphs of different size and composition that are generated as described in (\cite{BarrereSecMetric2019}). To create an AND/OR graph of size $n$, we first create the target node. Afterwards, we create a predecessor which has one of the three types (atomic, AND, OR) according to a probability given by a compositional configuration predefined for the experiment. For example, a configuration of $(60,20,20)$ means 60\% of atomic nodes, 20\% of AND nodes and 20\% of OR nodes. We repeat this process creating children on the respective nodes until we approximate the desired size of the graph~$n$. 

\subsection{Independent security measures}
\begin{figure}[!t]
	\centering
    \includegraphics[scale=0.65]{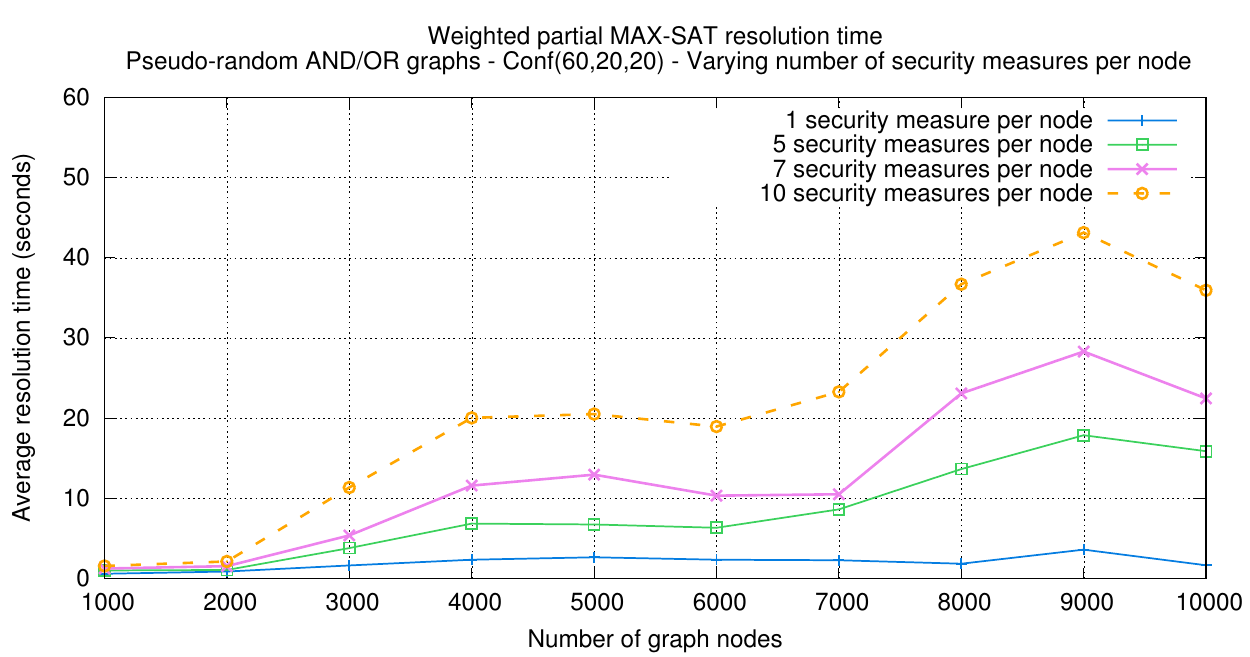}
    \caption{Scalability evaluation while increasing graph size}
    \vspace{-0.3cm}
	\label{fig:exp1}
\end{figure}

Our first set of experiments studies the impact on scalability and performance when we increase the number of security measures applied independently on each network node. Figure~\ref{fig:exp1} shows the results of this evaluation over AND/OR graphs with up to 10000 nodes. 

We have measured the MAX-SAT resolution time for graphs of different sizes in four sub-experiments that use a different number of independent security controls (1, 5, 7 and 10) on each graph node. Each sub-experiment has been repeated 10 times and we have taken the average results. As expected, we can observe that the more security measures we use to protect each node independently, the more time is required to compute the underlying security metric. As explained in (\cite{BarrereSecMetric2019}), even when there are small time variations on each sub-experiment due to the compositional characteristics of some random AND/OR graphs, the overall behaviour remains relatively stable. 

\begin{figure}[t]
    \centering
    \includegraphics[scale=0.65]{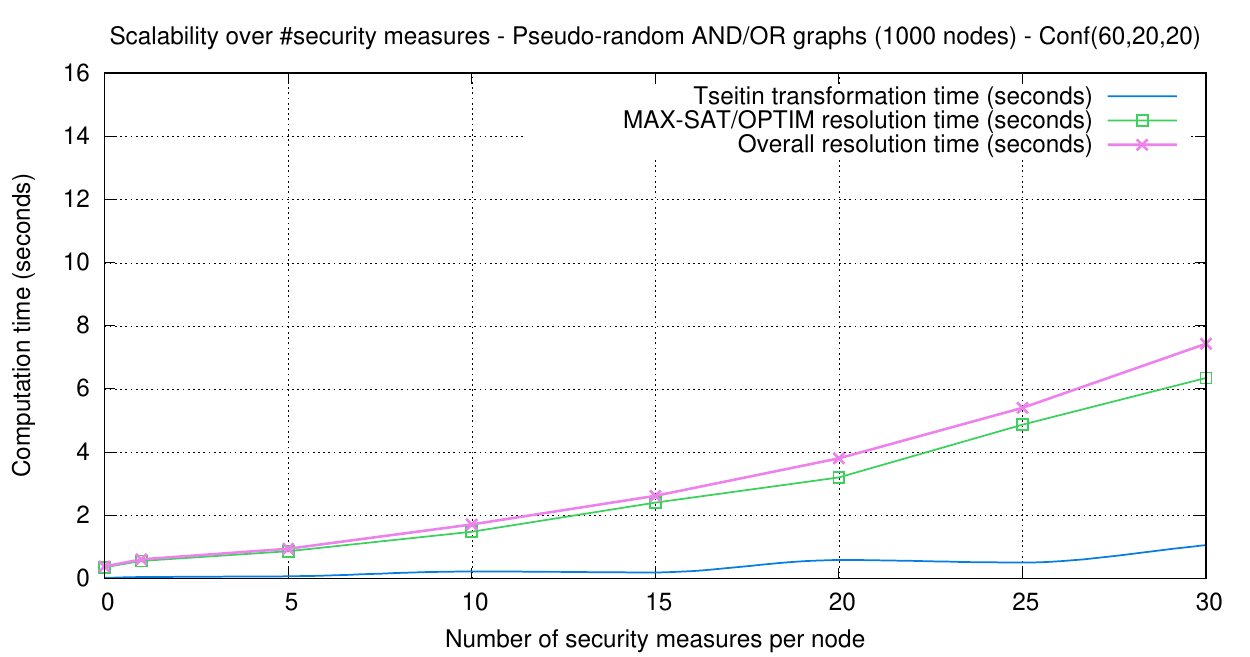}
    \caption{Performance while increasing nodes' measures}
    \vspace{-0.3cm}
    \label{fig:exp2}
\end{figure}

Figure \ref{fig:exp2} shows a closer look at the logical transformation and MAX-SAT resolution times for graphs with 1000 nodes while increasing the use of independent security measures. We can observe that the MAX-SAT resolution time grows polynomially. This is essentially explained by the fact that each node variable $n$ is replaced by a larger disjunction with $x$ logical variables (for $x$ security measures) plus the node variable itself. Because each node is protected by a different set of security measures (no overlapping), such replacement just increments the size of the formula by a factor of $x$. Therefore, the overall process is still solved in polynomial time. In hypergraph terms, the smaller the hyperedges, the lower the computation time. 

The previous experiments are focused on security measures that are applied individually on each ICS component. As mentioned before, however, many security measures may be used to protect two or more components altogether in practice, e.g. fenced areas. In the next section, we evaluate our approach considering multiple overlapping security measures.

\subsection{Overlapping security measures}

\begin{figure}[!t]
    \centering
    \includegraphics[scale=0.65]{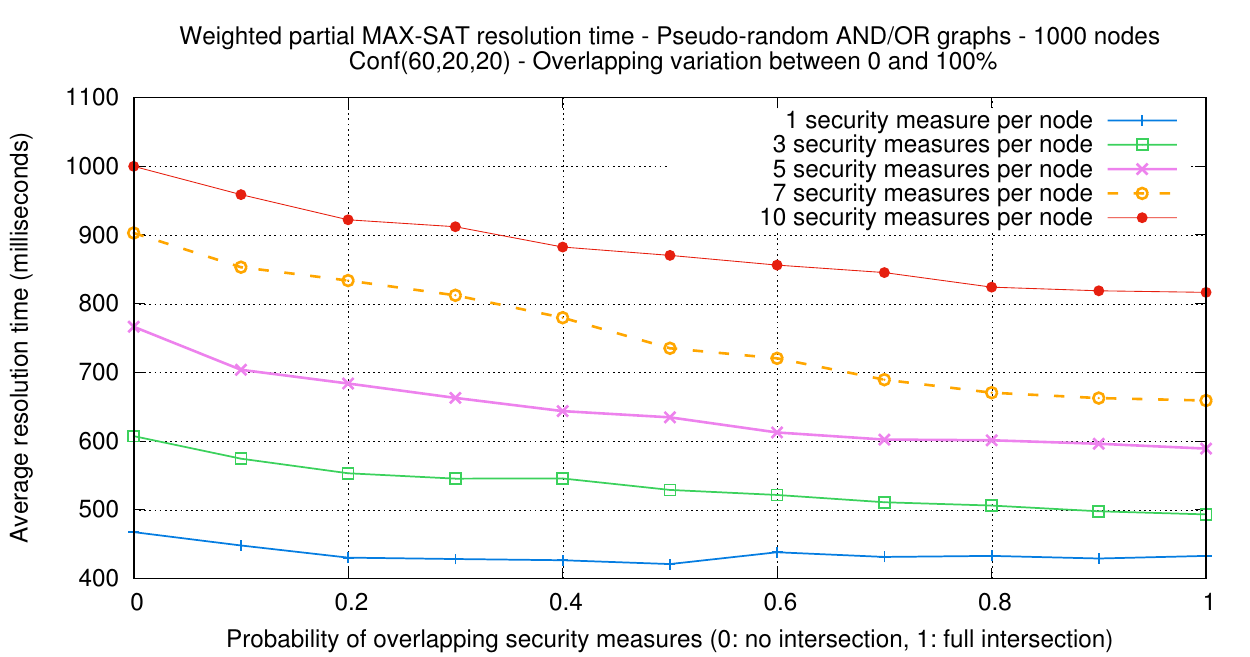}
    \caption{Variation analysis of overlapping measures}
    \vspace{-0.3cm}
    \label{fig:exp3}
\end{figure}

In order to analyse scenarios where two or more nodes may be protected by the same security control, we use a simple probabilistic method to generate a protection assignment as follows. Let $x$ be the number of security measures to be applied on each graph node $n \in V_{AT}$. We then traverse the set $V_{AT}$, and for each node, we stochastically choose whether to assign the same security measure used with the last node, or conversely, to use a new one. In mathematical terms, we apply the same security control with probability $p$ (positive overlapping), or we apply a new one with probability $1-p$ (no overlapping).  
We repeat the above procedure $x$ times. Figure~\ref{fig:exp3} shows the behaviour of the MAX-SAT resolution time over graphs with 1000 nodes that have been protected following the previous assignment. 

\begin{figure}[!b]
    \centering
    \includegraphics[scale=0.65]{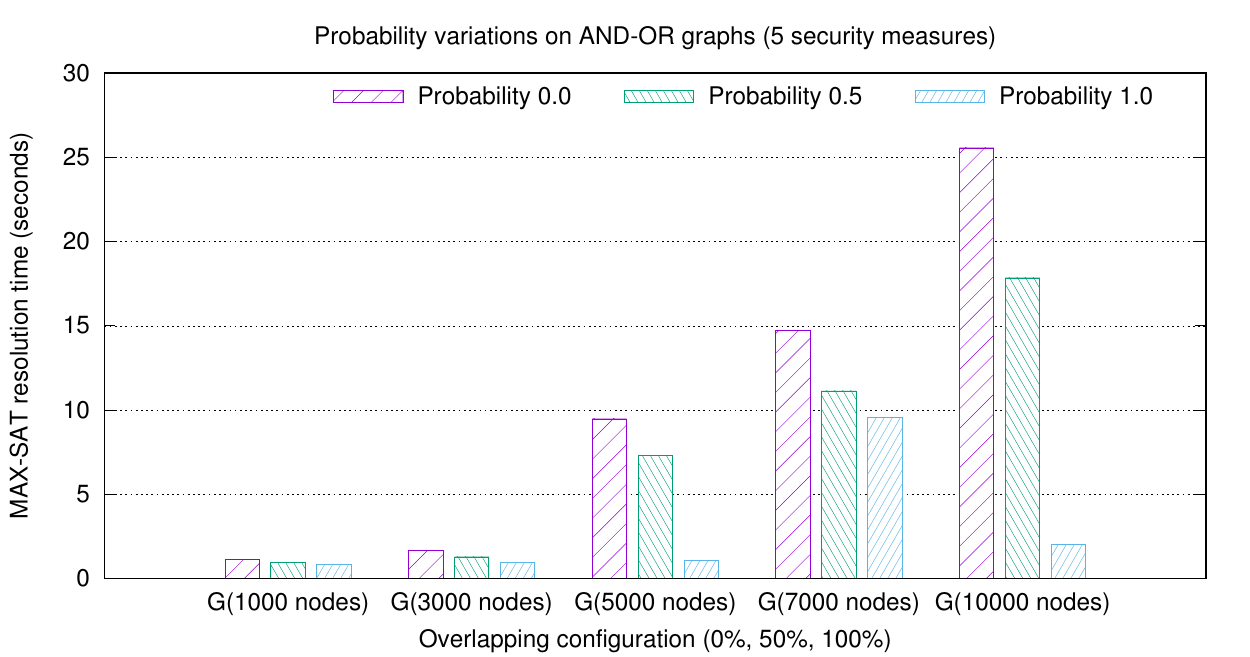}
    \caption{Overlapping analysis on graphs of different sizes}
    \vspace{-0.3cm}
    \label{fig:exp4}
\end{figure}

We can observe that, as the probability of overlapping increases from 0 to 1, the MAX-SAT resolution time decreases. In other words, the greater the level of overlapping, the easier is for the MAX-SAT solver engine to find the solution. In addition, this behaviour is observed independently of the number of security measures applied in the experiment. In logical terms, this happens because a security measure that protects many nodes will appear on the logical expansion of all of them (step~2, Section~\ref{sec:problem-spec}), and therefore, the MAX-SAT solver leverages such interdependency to speed up the overall resolution process. We have performed a similar analysis on larger AND/OR graphs and the results indicate the same behavioural pattern, as shown in Figure~\ref{fig:exp4}. 

The experiments involve AND/OR graphs with 1000 to 10000 nodes, using 5 security measures on each node. 
As expected, the results suggest that the more nodes are protected by the same security measures (i.e. higher probability), the faster is the resolution process. In the next section, we validate our approach through a comprehensive case study.

%% file: sections/case-study.tex
\section{Case study}
\label{sec:case_study}
 
Our case study is focused on water transport networks (WTNs) where we examine the applicability of our approach over real WTNs typically deployed in European countries.

\subsection{Case study description} 
\begin{figure}[!t]
 	\centering
 	\begin{minipage}{0.46\textwidth}
 		\centering 		
 		\includegraphics[scale=0.25]{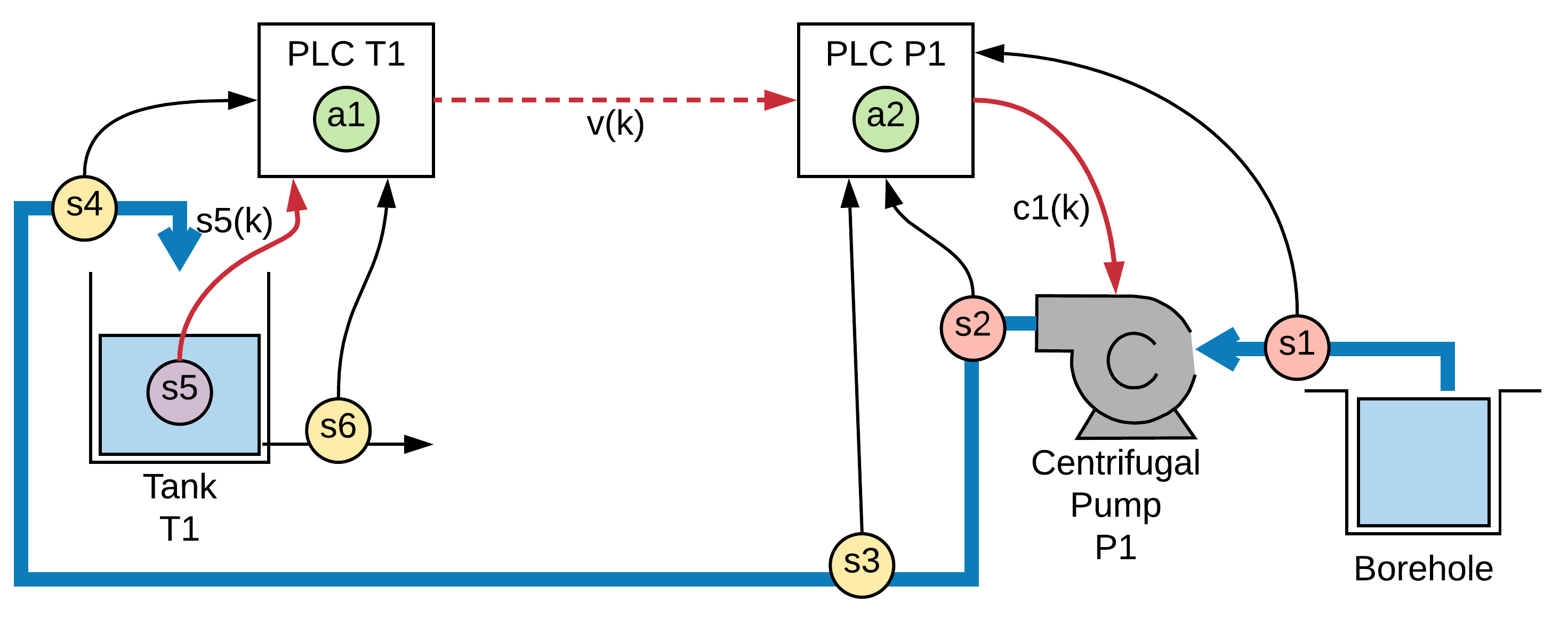}
        \caption{Basic WTN component (\cite{Nicolau2018})} 
		\vspace{-0.4cm}
 		\label{fig:case-study-1}
 	\end{minipage}\hfill
 \end{figure}

Typical WTNs are composed of the following main physical elements: (i) tanks, (ii) pumping stations, (iii) water sources (e.g., boreholes), and (iv) pipes. To monitor the status of each element, utilities deploy electronic sensing devices and collect measurements regarding the flow, pressure, level, and quality of the water that flows in the system. A typical configuration found in several water utilities (see \cite{Trifunovic2006}) is similar to the one shown in Figure \ref{fig:case-study-1}. The same structure appears repeatedly in larger infrastructures. In this work, we focus on the subsystem shown in \hbox{Figure \ref{fig:case-study-1}}. 

In this setup, drinking water is extracted from a water source (e.g., a borehole or another tank) using a pump. The pump increases the water pressure which pushes the water into a tank, which may be located a few kilometres away at a higher elevation. The water tank is then used to provide water to consumers, as well as to transfer water to other subsystems, for instance, through another pump-tank subsystem. 
 
The subsystem shown in Figure \ref{fig:case-study-1} involves the following sensing elements: a pressure sensor before the pump ($s1$), a pressure sensor after the pump ($s2$), and a water flow sensor ($s3$) measuring the pump outflow. At the water tank, flow sensors ($s4$, $s6$) may also be installed for monitoring the inflow and outflow respectively. For its operation, the control system is comprised of two Programmable Logic Controllers (PLCs); one situated at the pump and the other at the water tank. These PLCs are connected to the system's sensors and actuators, and execute programs to achieve the control objectives. More specifically, the sensing node $s5$ provides the water level state measurement $s5(k)$ to the agent $a1$ in PLC-T1, where $k$ is the discrete time step. Then, the control logic is executed, and the result $v(k)$ is transmitted to PLC-P1, where another control logic $a2$ is executed. Agent $a2$ instructs the contactor (i.e., an electrically operated relay) through a signal $c1(k)$ to turn on/off the pump, should the pump flow $s3$ be below a threshold. 
 
 \subsection{Data collection and preparation}
 
Various security measures are applied by water utilities in order to protect the components of their systems against malicious actors. We have acquired data from a number of water utilities and public information sources in order to: (i) determine typical measures used to protect their infrastructures, and (ii) identify components that are protected by multiple overlapping measures. 

Table \ref{tab:measures} presents a sample list of the measures acquired. We evaluate three different factors in order to calculate the cost of the attacker to compromise a security measure: (i) skills/knowledge required to design and execute the attack ($f1$), (ii) tools needed for the attack ($f2$), and (iii) time needed to execute the attack($f3$). We use a three-point scale to rate the three factors for each measure, as shown in Table \ref{tab:rating}. 

{    \renewcommand{\arraystretch}{1.25} 
  \begin{table}[!h]
 	\martinTableFontsize
 	\centering 	
 	\begin{tabular}{|p{1.48cm}|p{1.68cm}|p{1.88cm}|p{1.64cm}|}
 		\hline
 		{\bf Factor / Rate} & {\bf 1} & {\bf 2} & {\bf 3}  \\
 		\hline
 		\hline
         {\bf Skills} ($f1$) & no special skills / knowledge & advanced skills / knowledge & expert skills / knowledge \\
 		\hline
 		{\bf Tools} ($f2$) & off-the-shelf tools & non-conventional tools required & specialized tools \\
 		\hline 		
         {\bf Time} ($f3$) & $\leq 10$ min & 10-30 min &  $\geq30$ min \\
 		\hline
 	\end{tabular}
 	\vspace{0.1cm}
 	\caption{Attacker's cost - Three-point rating scale}
 	\label{tab:rating}
 \end{table}
}

\vspace{-0.2cm}

Then, for each collected measure $m$, we calculate the \emph{attacker cost} $\costprime(m)$ as the product of each individual rating: $ \costprime(m) = f1\times f2\times f3$. 
  
The cost of each component determines the level of difficulty an attacker will have to compromise it. The security measures  along with their individual ratings and attack costs are depicted in Table \ref{tab:measures}. 
 
{    \renewcommand{\arraystretch}{1.25}  
 \begin{table}[!h]
 	\martinTableFontsize
 	\centering
 	\begin{tabular}{|p{0.9cm}|p{0.6cm}|p{0.6cm}|p{0.6cm}|p{0.6cm}|p{2.6cm}|}
 		\hline  		
 		{\bf Measure} & {\bf Skills} & {\bf Tools} & {\bf Time} & {\bf Attack cost} & {\bf Description} \\
 		\hline
 		\hline
 		F1 & 1 & 1 & 1 & \textbf{1} & Fenced area (wire)\\
 		\hline
 		F2 & 1 &  2 & 1 & \textbf{2} & Fenced area (locked underground facility)\\
 		\hline
 		B1 & 1 &  1 & 2 & \textbf{2} & Building + regular lock \\
 		\hline
 		B2 & 2 &  2 & 2 & \textbf{8} & Building + secure lock \\
 		\hline
 		A1 & 2 &  3 & 2 & \textbf{12} & Door alarm \\
 		\hline
 		A2 & 3 &  2 & 3 & \textbf{18} & Alarm on telemetry box \\
 		\hline
 		A3 & 1 &  1 & 3 & \textbf{3} & Patrol unit \\
      	\hline
 		P1 & 1 &  2 & 1 & \textbf{2} & Locked box \\
 		\hline
 		P2 & 2 &  2 & 2 & \textbf{8} & Cable protection \\
 		\hline
 	\end{tabular}
 	\vspace{0.1cm}
     \caption{Typical security measures and attack costs}
 	\label{tab:measures}
 	\vspace{-0.36cm}
 \end{table}
}
 
\vspace{-0.2cm}
Based on this information, we have used our methodology to determine the security level of such infrastructures.

\subsection{Base WTN subsystem (no redundancy)}
According to the collected data, the base WTN subsystem shown in Figure \ref{fig:case-study-1} involves multiple security measures that simultaneously protect various components as shown in Table \ref{tab:scores-simple-scenario}. For example, agent $a1$ is protected by a wired fence (F1-2), located inside a building with a security lock (B1-1), and an alarm system (A3-1). Sensor $s5$ is also protected by the same measure instances but also by a protection box (P2-2). In order to make the scenario even more interesting, we assume the special case where $c1$ has been heavily protected and cannot be compromised (infinite cost). 
\begin{table}[!h]
    \centering  
    {\scriptsize        
        {    \renewcommand{\arraystretch}{1.25}
        \begin{tabular}{|p{1.6cm}|p{3.0cm}|p{2.2cm}|}
            \hline            
            {\bf Components} & { \bf Security measures} & { \bf Total cost} \\
            \hline
            \hline
            $s3$ & \{F2-1, P1-2, A2-2\}  & 22 \\          
            \hline
            $s5$ & \{F1-2, B1-1, A3-1, P2-2\} & 14\\
            \hline
            $a1$ & \{F1-2, B1-1, A3-1\} & 6 \\
            \hline
            $a2$ & \{F1-1, B2-1, P1-1, A2-1\} & 29 \\
            \hline
            $c1$ & \{F1-1, B2-1\} & 9 + inf (special case) \\
            \hline
        \end{tabular}
        }   
    }
    \vspace{0.1cm}
    \caption{Measures per component (base subsystem)}
    \label{tab:scores-simple-scenario}
    \vspace{-0.36cm}
\end{table}

\begin{figure}[!h]   
    \centering
    \hspace{-0.5cm}
    \subfloat[AND/OR graph]{\includegraphics[width=0.23\textwidth]{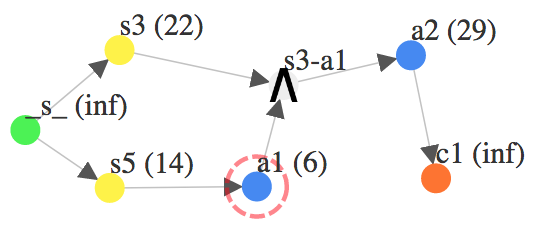}\label{fig:base-scenario-graph}}
    \hfill
    \subfloat[AND/OR hypergraph]{\includegraphics[width=0.23\textwidth]{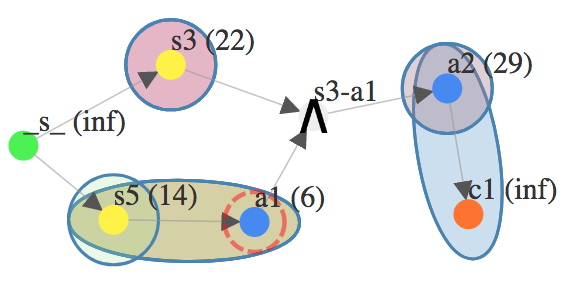}\label{fig:base-scenario-hypergraph}}
    \caption{Base scenario}
    \label{fig:base-scenario}
\end{figure}

\vspace{-0.4cm}
The total cost to compromise a component $n$ is computed as $\sum_{m\in S_n}\costprime(m)$, where $S_n$ is the set of security measures protecting $n$. Given the AND/OR specification of the base subsystem with no redundancy, we have run \tool in order to identify the set of critical ICS components and security measures, as shown in Figure \ref{fig:base-scenario}. 

\vspace{-0.4cm}

\vspace{-0.1cm}
Figure \ref{fig:base-scenario-graph} shows the AND/OR graph of the network where, given the applied measures, \tool has identified agent $a1$ at PLC-T1 as the weakest point that can disable actuator $c1$. Its compromise implies to bypass three security measures (F1-2, B1-1, \hbox{A3-1}) with a total cost of $6$. Figure \ref{fig:base-scenario-hypergraph} shows the AND/OR hypergraph of the system involving its multiple overlapping measures. Agent $a1$ is responsible for measuring the water level of the tank and deciding whether to send a signal to turn on/off the pump. Note that sensor $s5$, which also measures the level of the tank, was not identified as a critical node as it is guarded with stronger security measures and a total attack cost of $14$ (see Table~\ref{tab:scores-simple-scenario}).

\subsection{Extended WTN subsystem with redundancy}

WTN systems are typically set up using the minimum configuration. However, additional sensors and agents can be used to introduce analytical redundancy in order to ensure the reliable operation of the system. In that context, we have analysed an extended scenario, detailed in (\cite{Nicolau2018}), involving the components and security measures listed in Table \ref{tab:scores}. Table \ref{tab:measure_cost} on the other hand shows the components protected by each measure instance and their costs. 

 \begin{table}[!t]
    \centering
    {\scriptsize
        {    \renewcommand{\arraystretch}{1.25}             
            \begin{tabular}{|p{1.6cm}|p{3.0cm}|p{2.2cm}|}
                \hline                
                {\bf Components} & { \bf Security measures} & { \bf Total cost} \\
                \hline
                \hline
                $a2, a7, a8, a10$ & \{F1-1, B2-1, P1-1, A2-1\} & 29 \\
                \hline
                $a1, a3, a9$ & \{F1-2, B1-1, A3-1\} & 6 \\
                \hline
                $s1, s2$ & \{F1-1, B2-1\} & 9 \\
                \hline
                $c1$ & \{F1-1, B2-1\} & 9 + inf (special case) \\
                \hline
                $s3$ & \{F2-1, P1-2, A2-2\}  & 22 \\
                \hline
                $s4$ & \{F1-2, B1-1, A3-1, P2-1\} & 14 \\
                \hline
                $s5$ & \{F1-2, B1-1, A3-1, P2-2\} & 14\\
                \hline
                $s6$ & \{F2-2, P1-3, A2-3, A3-1\} & 25 \\
                \hline		
            \end{tabular}
        }
    }
    \vspace{0.1cm}
    \caption{Measures per component (redundant subsystem)}
    \label{tab:scores}
    \vspace{-0.66cm}
\end{table}

\begin{table}[!h]
	\centering
	{\scriptsize
        {    \renewcommand{\arraystretch}{1.25} 		
		\begin{tabular}{|p{1cm}|p{1.0cm}|p{1.0cm}|p{3.4cm}|}
			\hline			
            {\bf Measure instance} & { \bf Measure type} & {\bf Attacker cost} & {\bf Protection range} \\
			\hline
			\hline
			F1-1 & F1 & 1 & $\{a2, a7, a8, a10, c1, s1, s2\}$ \\ 
			\hline
			F1-2 & F1 & 1 & $\{a1, a3, a9, s4, s5\}$ \\ 
			\hline
			F2-1 & F2 & 2 & $\{s3\}$ \\ 
			\hline
			F2-2 & F2 & 2 & $\{s6\}$ \\ 
			\hline
			B1-1 & B1 & 2 & $\{a1, a3, a9, s4, s5\}$ \\ 
			\hline
			B2-1 & B2 & 8 & $\{a2, a7, a8, a10, c1, s1, s2\}$ \\ 
			\hline
			A2-1 & A2 & 18 & $\{a2, a7, a8, a10\}$ \\ 
			\hline
			A2-2 & A2 & 18 & $\{s3\}$ \\ 
			\hline
			A2-3 & A2 & 18 & $\{s6\}$ \\ 
			\hline
			A3-1 & A3 & 3 & $\{a1, a3, a9, s4, s5, s6\}$ \\ 
			\hline
			P1-1 & P1 & 2 & $\{a2, a7, a8, a10\}$ \\ 
			\hline
			P1-2 & P1 & 2 & $\{s3\}$ \\ 
			\hline
			P1-3 & P1 & 2 & $\{s6\}$ \\ 
			\hline
			P2-1 & P2 & 8 & $\{s4\}$ \\ 
			\hline
			P2-2 & P2 & 8 & $\{s5\}$ \\ 
			\hline
		\end{tabular}
        }
	}
	\vspace{0.1cm}
	\caption{Components per measure instance}
	\label{tab:measure_cost}
	\vspace{-0.66cm}
\end{table}

\begin{figure}[!b]
		\centering		
        \includegraphics[scale=0.44]{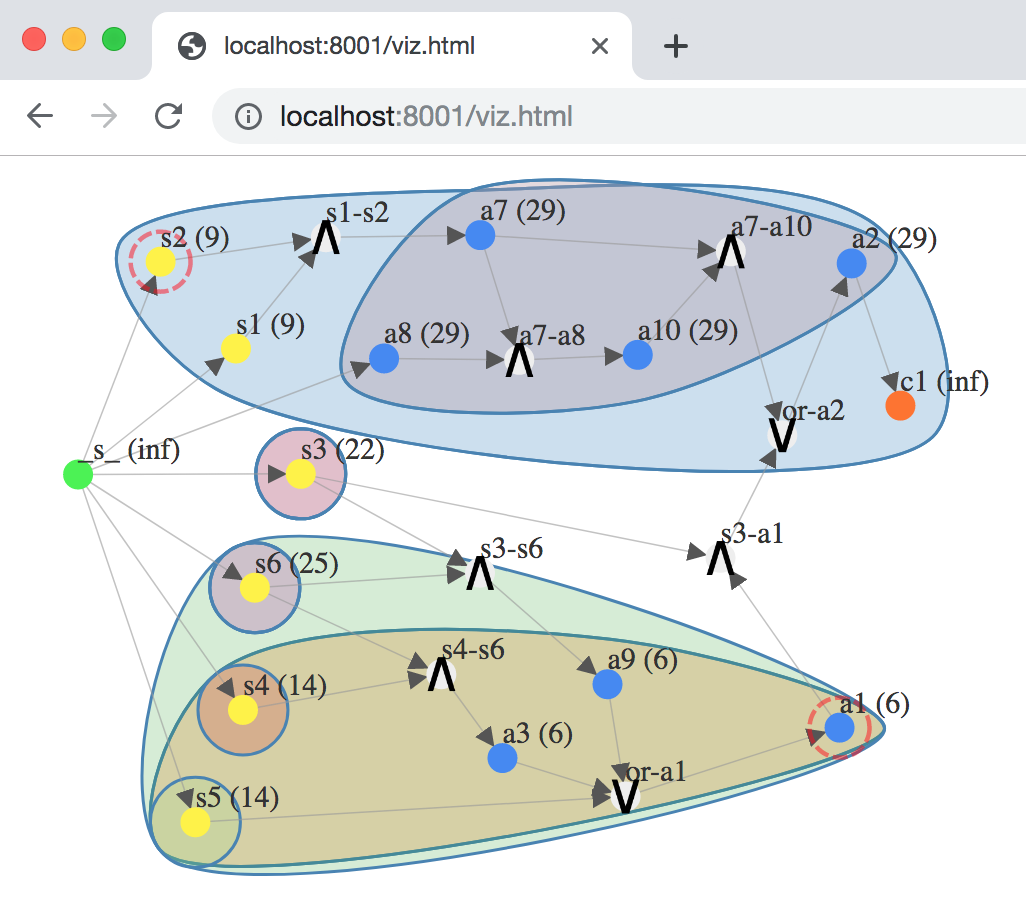}
		\caption{\hbox{AND/OR hypergraph with overlapping measures}}        
       	\vspace{-0.86cm}
		\label{fig:scenario-expanded-hypergraph}
\end{figure}

The structure of the network as well as the critical nodes identified by \tool are shown in Figure~\ref{fig:scenario-expanded-hypergraph}. The optimal strategy indicated by the tool involves agent $a1$ and sensor $s2$ as the critical nodes and five different measure instances (F1-2, B1-1, \hbox{A3-1}, F1-1, B2-1) that should be violated so as to disable actuator $c1$, with a total attack cost of $15$. Note that the security level of this configuration is much higher than the settings without redundancy. 

It is important to note that, as opposed to the base scenario, identifying critical components and security measures on larger scenarios with various components and multiple overlapping security measures becomes significantly harder. In that context, our approach provides strong support for security decision-making, prioritising mitigation plans, and increasing the resilience of ICS environments.  
 

%% file: sections/related-work.tex
\vspace{-0.2cm}
\section{Related work}
\label{sec:rw}

Since the early 2000s, many research efforts have been produced to understand and improve the security of industrial control systems and critical national infrastructure (\cite{Desmedt2004}). These works have inspired the need for taking into account the cyber-physical dependencies between ICS components and being able to combine them in order to provide quantifiable measurements (\cite{Humayed2017}). As such, our approach builds upon the contributions presented in (\cite{Nicolau2018}, \cite{BarrereSecMetric2019}). The latter provides a complete AND/OR graph-based modelling capable of grasping complex interdependencies among CPS components. However, the approach only covers security controls that are applied independently to each ICS component, thus involving a single score for each one of them. In this paper, we extend such an approach by allowing ICS components to share multiple overlapping cyber-physical security measures and providing an overall security score for the ICS network. 

From a graph-theoretical perspective, the underlying base security metric used in this paper, presented in (\cite{BarrereSecMetric2019}), looks for a minimal weighted vertex cut in AND/OR graphs. \hbox{This is an} \hbox{NP-complete} problem as shown in (\cite{Desmedt2004, Jakimoski2004, Souza2013}). While well-known algorithms such as Max-flow Min-cut (\cite{Ford1962}) and variants of it could be used to estimate such metric over OR graphs in polynomial time, their use for general AND/OR graphs is not evident nor trivial as they may fail to capture the underlying logical semantics of the graph. \hbox{In that} context, we leverage \hbox{state-of-the-art techniques which} excel in the domain of logical satisfiability and boolean optimisation problems (\cite{Davies2011}). 

A close research area to our problem includes the domain of attack graphs (\cite{Wang2017}, \cite{Barrere:CNSM2017}, \cite{Shandilya2014}). While attack graphs are mainly focused on depicting the many ways in which an attacker may compromise assets in a computer network, our approach is essentially different as we consider that network nodes can be equally compromised. In addition, attack graphs usually take into account only cyber lateral movements, without considering operational cyber-physical dependencies among components (\cite{Humayed2017}). Moreover, real ICS models based on AND/OR graphs might also be cyclic, thus presenting the interdiction problem (\cite{Altner2010}). We deal with cycles using a similar approach to that considered in (\cite{Wang2017}). 

Other attempts to identify critical cyber-physical components have been made in the domain of network centrality measurements (\cite{Deng2018}). While useful in many types of scenarios (\cite{Steiner2018}), almost all of them are focused on OR-only graph-based models for IT networks. In addition, we realise that automating ICS asset mapping is not an easy task for different reasons, among these, because active probing and scanning may be too intrusive, which might raise concerns about operational disruptions. However, this is a premise that many security platforms already take into account, e.g., in the form of passive monitoring (\cite{CyberXReport2019}). 

%% file: sections/conclusions.tex
\vspace{0.1cm}
\section{Conclusions and future work}
\label{sec:conclusions-fw}

Industrial control systems typically involve a large spectrum of overlapping cyber-physical security measures used to protect their operational components. As such, understanding which security measures and ICS components should be compromised so as to disturb the normal operation of the system with minimal cost for an attacker is a challenging task. In this paper, we solve this problem via an efficient mechanism based on AND/OR hypergraphs, which is able to capture complex interdependencies among ICS components and the measures used to protect them. Our approach extends the MAX-SAT-based techniques presented in (\cite{BarrereSecMetric2019}) and is able to scale to thousands of nodes in seconds, as demonstrated in our experimental evaluation. In addition, we have presented a thorough case study conducted over a realistic water transport network that shows the applicability of our method. 

As future work, we plan to further analyse our approach on other classes of ICS systems such as smart grids and power plants. We also plan to extend our methodology to integrate attack graphs at the cyber level, socio-technical aspects, multi-target attacks, and defence budget constraints. Redundant components sometimes handle only a fraction of the functions provided by main components. We plan to refine our model to cover this aspect as well as standard fault-tolerant techniques such as triple modular redundancy (TMR)~(\cite{Kastensmidt2005}). Finally, we aim at further investigating automated mechanisms to generate AND/OR graph-based models for ICS environments.